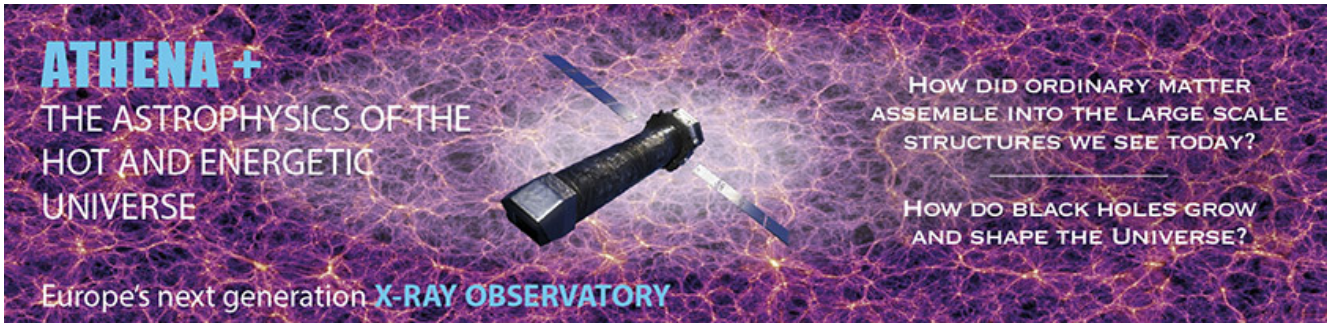

# The Hot and Energetic Universe

An *Athena+* supporting paper

## End points of stellar evolution

Authors and contributors

**Christian Motch, Jörn Wilms,** Didier Barret, Werner Becker, Slavko Bogdanov, Laurence Boirin, Stéphane Corbel, Ed Cackett, Sergio Campana, Domitilla de Martino, Frank Haberl, Jean in 't Zand, Mariano Méndez, Roberto Mignani, Jon Miller, Marina Orio, Dimitrios Psaltis, Nanda Rea, Jérôme Rodriguez, Agata Rozanska, Axel Schwope, Andrew Steiner, Natalie Webb, Luca Zampieri, Silvia Zane

# 1. EXECUTIVE SUMMARY

The endpoints of stellar evolution, white dwarfs, neutron stars and black holes are key laboratories to study matter in its most extreme conditions. High compactness, strong gravity and extreme magnetic fields turn these objects into unrivalled test beds for nuclear and particle physics, general relativity, and extreme magneto-hydrodynamical phenomena.

Accretion is a fundamental process that powers many types of astrophysical sources, from proto-stars, black holes in active galaxies, to gamma-ray bursts. Accreting white dwarfs, neutron stars, and stellar-mass black holes in X-ray binary systems are among the brightest sources in the X-ray sky. They are unique nearby laboratories to study the dependency of accretion physics on accretion rates and magnetic field conditions. X-ray observations of accreting white dwarfs allow us to study the behaviour of matter in very strong magnetic fields of up to $10^8$ G, neutron stars allow the study of matter under the most extreme conditions of density and magnetic field strength (up to $10^{14}$ G), while black hole X-ray binaries allow us to test General Relativity in the strong field limit - accreting black holes are the only sources where we are able to study directly phenomena occurring within a few gravitational radii from the black hole. Black hole binaries from their quiescent and very sub-Eddington stages up to their super-Eddington manifestations as ultra-luminous X-ray sources offer the wide range of accretion rates required. As such, these binaries serve as scaled down models of what happens close to the super-massive black holes found in the centres of most galaxies, albeit with a much shorter dynamical time, and allow us to study accretion flows on dynamical time scales inaccessible in any super-massive black hole.

Neutron stars are the most compact objects whose surface is directly accessible to astrophysical observations. Understanding their properties requires confronting observations with fundamental physics in conditions not reproducible in ground-based laboratories. The unprecedented throughput and spectral resolution of *Athena+* will allow new diagnostic studies of these sources even at very faint flux levels. Multi-wavelength observations with *Athena+* and contemporaneous large facilities in other wavebands will provide crucial spectral and timing information on the emission mechanisms operating in rotation powered pulsars. Most importantly, in the physical conditions prevailing inside neutron stars, the way nucleons interact remains uncertain. Accurate measurements of radius and mass are crucial to constrain the equation of state and determine the state of matter in neutron star cores. This will in turn provide very important complementary information to terrestrial experiments, which cannot probe this high density and low temperature regime. The sensitivity, high spectral resolution, and high count rate capability of *Athena+* will provide multiple, independent measurements of the neutron star mass/radius relation in a wide range of environments and conditions.

# 2. OBSERVATIONAL AND INSTRUMENTAL CONTEXT

For many of the brightest and therefore potentially most interesting sources, limitations of current instruments significantly hamper our ability to make progress in our understanding of these sources as most of the flux is discarded for technical reasons. The relevant metric is the product of the effective area, $A_{eff}$, and the fraction of the time that a detector can register events, $F_{live}$. In current instruments, $F_{live}$ can be as small as a few percent, since a large fraction of events has to be discarded due to telemetry limitations and due to deadtime in the detector electronics. For example, for a 0.2 Crab source ($4·10^{-9}$ erg cm$^{-2}$ s$^{-1}$), in the diagnostically important iron band at 6.4 keV, both *XMM-Newton* and *Chandra* in their fastest modes yield $A_{eff}·F_{live}$ ~ 30 cm$^2$. *Astro-H*'s calorimeter will have $A_{eff}·F_{live}$ = 250 cm$^2$. In contrast, for the *Athena+* WFI and X-IFU, $A_{eff}·F_{live}$ = 2500 cm$^2$. Note that because of strong source variability, in many cases longer observations *cannot* be used to compensate for the shortcomings of current detectors. The low livetime of, e.g., the EPIC-pn in its burst mode (3%), means that in order to accumulate a spectrum with 10 ks exposure the duration of the observation must be ~300 ks. This means that any fast variability cannot be resolved. The *Athena+* WFI, on the other hand, would accumulate a similar quality spectrum in a few ks. *Every spectrum of a bright source with Athena+ will be an order of magnitude better than spectra obtained with earlier facilities.*

In addition, simultaneous multi-wavelength observations are necessary for modelling the spectral energy distributions of compact objects from the radio or optical up to the X-ray and TeV gamma-ray bands. *Athena+* will operate at a time when radio and optical monitoring will be routinely available through *SKA* and *LSST*. The high sensitivity of these instruments opens a new window of discovery not based on the transient X-ray sky (as was the case for the past 20 years) but on the transient radio or optical skies. This opens the possibility, for example, to study the onset of accretion onto black holes and the subsequent formation of radio jets, one of the most important and unsolved problems in astrophysics.

In the following, we describe three areas where *Athena+* will make significant progress in studies of such objects.

## 3. GALACTIC BLACK HOLES AND MICROQUASARS

### 3.1. Introduction

The inflows of accretion disks are, in most sources, systematically paired with powerful ejections or jets. While jets from super-massive black holes in quasars have been known since the early 1960s, that the radio emission of their Galactic scaled-down versions associated with jets was only realized in 1992 with the discovery of "microquasars" (Mirabel & Rodriguez, 1992). Only recently has it been realised that accretion and jets are linked in a non-random and non-trivial way. 'Low' source X-ray states are associated with powerful compact (AU scale) continuous ejections, while 'high' source X-ray states show no jets. The transition between states is accompanied by discrete ejections, which are linked to relativistic, and sometimes superluminal, outflows. Jets carry a large amount of both accretion energy and material. The jet power can be comparable to the bolometric luminosity of the system. Jets are usually observed from the radio to the infrared, with possible evidence for emission up to the soft gamma-ray range (Laurent et al., 2011). Because of the several order of magnitude strong accretion rate variations observable in these sources and the large dynamic range of observed accretion rates, these objects allow us to study accretion physics over the extremely wide range from sub-luminous to super Eddington accretion rates through broad band spectroscopic studies.

The physics responsible for the different modes of accretion and the formation and powering of all types of jets is largely not understood. Some of the currently debated questions concern the formation and destruction of jets and their relation to the accretion disc behaviour; the origin of all X-ray emitting media; the evolution and role of the accretion disc and accretion rate during an outburst: is the disc truncated, or does a cold disc exist down to the last stable orbit during low states? How do the sources behave in their off/quiescent states? Residual faint X-ray emission has been seen, e.g., with *Chandra*, but the sensitivity of current instruments did not permit the processes to be understood. How fast are transitions and what drives them? Are they always associated with jets? Because the accretion flows in active galactic nuclei and in microquasars are self-similar, observations of microquasars also yield insight into the behaviour of super-massive black holes. However, since the characteristic time scales in these systems scale with the mass of the compact object, these object classes allow complementary views on the same system: while microquasars allow us to study the evolution of the accretion flow on time scales that cannot be observed in active galaxies, active galaxies allow us to study the state of the flow "frozen in time".

### 3.2. Relativistic Disk Lines and Black Hole Spin: Order of Magnitude Improvements

Gravity in the strong field limit can only be probed close to spinning black holes. It is in this unique regime that orbital radii are no longer large compared to the Schwarzschild radius. Even double radio pulsars that decay via gravitational radiation cannot probe this limit. Iron emission lines produced in disks of gas orbiting black holes bear the imprints of their extreme conditions. They are shaped by special relativity owing to the velocities in their orbits, and by General Relativity as the orbits extend close to the black hole. Exactly how closely stable orbits can survive is set by the *spin* of the black hole. "Relativistic" iron lines, then, are our single best probes of the strong gravitational environment, and of black hole spin (e.g., Miller, 2007). Stellar-mass black holes in the Milky Way are local analogues of the massive black holes that power Seyfert AGN and quasars. Their proximity means that we observe a much higher flux, and can, in principle, obtain excellent spin constraints from relativistic lines (see Fig. 1). Such observations will translate into vastly improved spin measurements that can be used for the purposes of studying the black hole birth events (GRBs and/or SNe) that set stellar-mass black hole spins (e.g., Miller et al., 2011), and to study the relationship between black hole spin and relativistic jets (e.g., Russell et al., 2013). Due to the good low energy sensitivity of Athena+, independent confirmation of these spin measurements will come from modelling the accretion disc continuum of these sources in their soft states (e.g., Steiner et al., 2012, and therein) and through time-lag reverberation mapping (e.g., Uttley et al., 2011). See Dovciak, Matt, et al, 2013, *Athena+* supporting paper for a further discussion of these methods.

Athena+ will be sensitive enough to measure high signal to noise spectra from stellar-mass black holes also in nearby galaxies. Studying their relativistic iron lines and continua, Athena+ will deliver an order of magnitude more stellar-mass black hole spin measurements than any prior mission. *Whereas we currently have about 10 spin constraints in Galactic binaries, the Athena+ mission will increase this number by a factor of 10 due to its significantly larger collecting area (note that ~100000 photons are required to detect relativistic lines, e.g., Guainazzi et al., 2006).* This larger number of spin measurements allows us to study the distribution of spin values of black holes, which is related to the supernova mechanism forming them as well as to their accretion history.

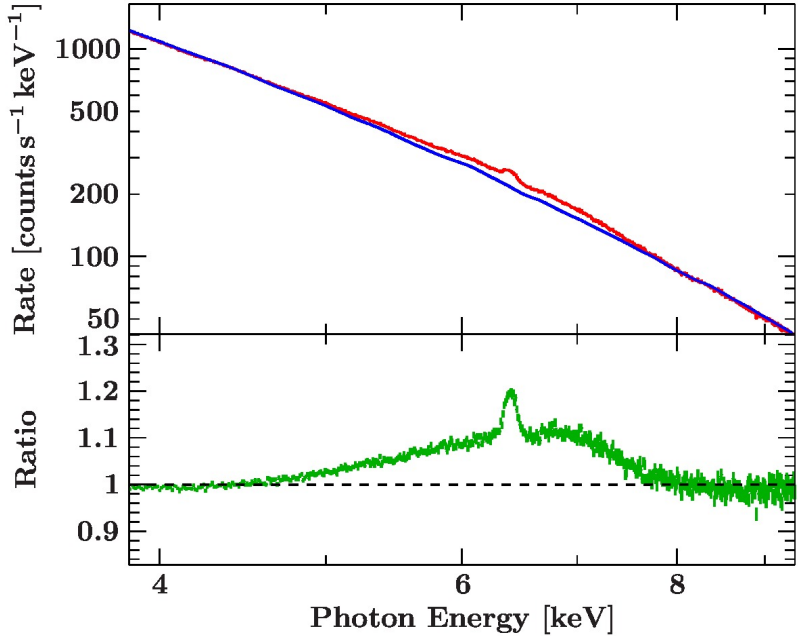

Figure 1: Simulation of a 10 ks Athena+ WFI observation of the black hole binary Cyg X-1 (based on XMM-Newton observations by Duro et al., 2011). Top: predicted count rate (red) and best fit model after setting the iron line flux to zero (blue). Bottom: ratio between the data and the best fit model with a zero iron flux. The ratio shows the iron line profile, which consists of a skew symmetric relativistic line originating close to the black hole, and a narrow core from material further away from the black hole. Even on such a bright (250 mCrab) object, current instrumentation can obtain similar high quality spectra only in much (>20x) longer exposures, where systematics of continuum and $N_H$ variation become important and dominate the uncertainty of the spin and continuum shape measurement.

## 3.3. High Resolution Spectroscopy of Accretion Flows Around Compact Objects

In addition to jets, many Galactic X-ray binaries exhibit outflows in which a significant fraction of the accreted materials is driven away from the accretion disk and back into the interstellar medium (e.g., Brandt & Schulz, 2000; Miller et al., 2008). The amount of mass lost in this wind can be close to the mass accretion rate and in black hole binaries the interaction between the disk wind and the jet might quench the jet (Neilsen & Lee, 2009). Disk winds are important because they serve to enrich the interstellar medium and because they can influence the overall energetics of the accreting system (Diaz Trigo & Boirin, 2012). They are violently variable on timescales of hours. Similar winds have also been seen in several AGN (e.g., Dauser et al., 2012). Their mass loss rates are still unknown, but if they are of similar power as the winds in Galactic sources they would have significant impact on the centres of active galaxies (See Cappi, Done, et al, 2013, *Athena+* supporting paper).

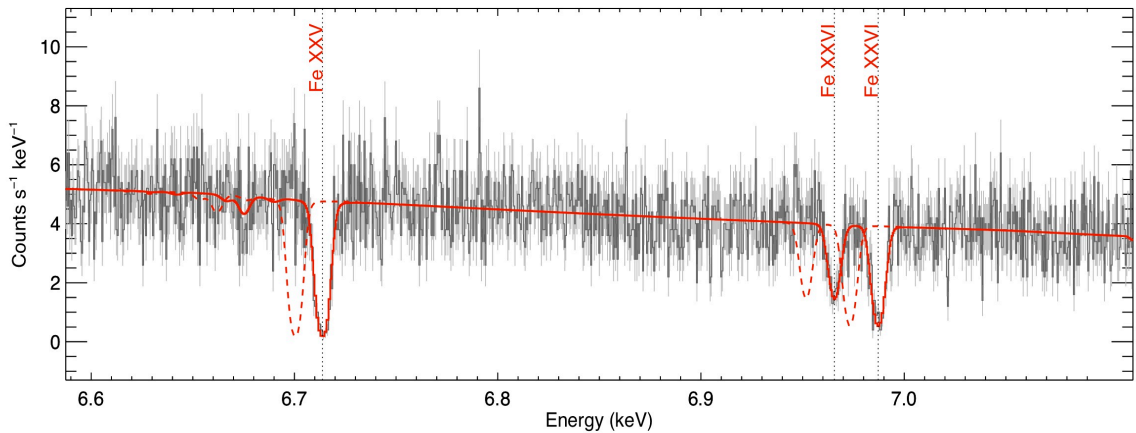

Figure 2: X-IFU simulation (10 ks exposure) of an outflow with a velocity of 600 km s$^{-1}$ in a typical LMXB, based on *XMM-Newton* data (Boirin et al., 2005; $F_{0.5-10keV}=2\cdot10^{-10}$ erg cm$^{-2}$ s$^{-1}$). The spectral continuum (a black body with kT=1 keV and a power law with a photon index of 1.9) is affected by absorption from a highly ionized wind ($N_H=3.6\cdot10^{22}$ cm$^{-2}$, ionization parameter log ξ=3.9) with a velocity of 600 km s$^{-1}$. The ionizing flux is described by a cutoff powerlaw with a cutoff energy of 44 keV and a photon index of 1.9. The red dashed line shows the zero velocity model while the solid line shows the fitted blue shifted model.

Mainly because of the low throughput and/or the low spectral resolution of current instruments, disk winds have only been studied in a small number of systems and therefore their origin is still debated. Are they radiation pressure driven (mainly through UV lines), thermodynamic winds (photoionization heating the disk surface to the point where the thermal speed of ions exceeds the local escape velocity), or of magnetohydrodynamic origin? To answer these questions, observations with high spectral resolution and large throughput are required to allow us to measure the ionization structure and density of the medium (through He-like ions, e.g., Miller et al., 2008) on typical time scales of the wind variability. Such data will allow us to pinpoint the location where the wind is launched (Fig. 2). Current facilities yield such measurements only for a few of the brightest sources, and none are able to resolve the required short timescales. They therefore average over different dynamic states of the wind, such that no quantitative statements on the wind launching mechanism or its dependence on source luminosity are possible.

### 3.4. Probing the highest mass accretion rates: Ultraluminous X-ray Sources

While in Galactic sources *Athena+* will be able to probe accretion flows at the lowest mass accretion rates, at the other extreme in terms of accretion rate, Ultraluminous X-ray sources (ULXs) remain one of the most intriguing mysteries of astronomy. Their luminosities exceed the Eddington limit for the luminosity of stellar-mass black holes, yet their extra-nuclear locations rule out a central super-massive black hole. Hence we are left with a choice of exotic scenarios – either ULXs contain a hitherto unseen class of accreting intermediate-mass black holes (IMBHs), or they embody a new regime of accretion physics in which the accretion flow somehow appears to exceed the Eddington limit. While a combination of environment and counterparts currently supports the presence of super-Eddington accretion in most ULXs (e.g., Roberts, 2007, Gladstone et al., 2009), a number of the brightest sources show properties consistent with harbouring IMBHs (e.g., Farrell et al., 2009). Either case has cosmological implications: IMBHs may be the relics of the population of black holes that seeded the first super-massive objects, whilst physical processes allowing the accretion of material at super-Eddington rates appear necessary for the rapid appearance of the first generation of supemassive black holes in the early universe.

The key evidence for ULXs as super-Eddington accretors currently comes from their unusual X-ray characteristics, mainly an X-ray energy distribution exhibiting a spectral break at a few keV that is the likely signature of strong Comptonization (Gladstone et al. 2009; Heil et al. 2009) and an absence of typical sub-Eddington black hole states transitions (Kajava & Poutanen 2009; Feng & Kaaret 2009). These signatures are detected in the brightest ~10 objects and require a high number (> 10,000) of detected photons. Such data will be available with Athena+ from sources with L>10$^{38}$ erg s$^{-1}$ within a few Mpc. In M81 alone, there are 4 such sources (Tennant et al., 2001). In nearby (<20Mpc) galaxies there are ~100 ULXs with L>10$^{39}$ erg s$^{-1}$ known (Swartz et al., 2004). The broad PSF of *XMM-Newton* currently inhibits studies of them in the crowded inner regions of nearby galaxies, and *Astro-H* will be even worse, but the 5" PSF of Athena+ will allow the separation of the ULX emission from other sources.

Deep high resolution observations will allow us to determine the chemical abundances in the source environment from photo-ionization edges, only marginally attainable with present instrumentation (e.g., Winter et al., 2007), but a crucial piece of information for understanding the origin of these systems and their black holes (e.g., Zampieri & Roberts, 2009; Linden et al., 2010; Mapelli et al., 2010). These observations will also offer the opportunity to search for blue-shifted narrow absorption features from disk winds, a key prediction of super-Eddington models (e.g., Poutanen et al., 2007, Ohsuga & Mineshige 2011).

## 4. NEUTRON STARS

### 4.1. The equation of state

Arguably the most exotic physical conditions in the Universe are found in the cores of neutron stars (NSs). While Quantum Chromodynamics (QCD) has been tested in various ways on the earth, the limit of high densities and low temperatures QCD can only be tested in the extreme astrophysical environment of NS cores. Here QCD predicts a range of rich behaviours depending on assumptions about the way particles interact in this extreme regime. Exotic excitations such as hyperons, or Bose condensates of pions or kaons may appear. It has also been suggested that at very high densities there may be a phase transition to strange quark matter. Therefore, constraining the state of matter in neutron stars provides unique complementary information to terrestrial experiments. The key "observable" to distinguish between the various models is the Equation of State (EoS), which governs the mechanical equilibrium structure of bound stars and which, in astrophysical terms, can be translated into the mass-radius relationship of neutron stars. Definitive constraints can only be derived from the measurement of both masses and radii of individual neutron stars over a wide range of masses, in particular at higher masses where differences between theoretical EoS are the most marked (e.g., Steiner et al., 2013). Neutron stars in mass-transferring binaries, as well as their end products (recycled pulsars), will offer the wide range of masses suitable to address this fundamental issue. The X-rays generated around neutron stars will be used to constrain their masses and radii, using multiple redundant and complementary diagnostics in a variety of environments and conditions (e.g., isolated stars, binary radio pulsars, quiescent cooling systems, X-ray bursters and accreting binaries). These diagnostics, which require both high and medium spectral resolutions as well as high sensitivity and capability to observe at relatively high fluxes, are for the first time fully implemented in the WFI and X-IFU designs.

A promising way to constrain the mass-radius relation for neutron stars is through observations of thermal emission from transiently accreting NSs when they are in their quiescent state, especially those for which a reliable distance exists, e.g. for systems in globular clusters (Webb & Barret, 2007) or for those sources for which in the 2020s reliable parallaxes will have been measured with *GAIA*. The most up-to-date versions of the non-magnetic models assuming a hydrogen atmosphere have been shown to provide adequate fits to the quiescent X-ray spectra of all systems studied in detail. However, the current instrumentation does not allow us to narrow down enough the allowed EoS. The gain in effective area offered by *Athena+* will allow for simultaneous constraints on both the mass and the radius with uncertainties that are small enough to exclude a large region of the *M-R* plane. In a similar manner, the cooling branches of X-ray bursts exhibiting photospheric radius expansion as a result of their Eddington luminosity allow us to accurately measure both mass and radii when their distance is known (e.g., Özel et al., 2010). The shape of the thermal pulse of rotation-powered millisecond pulsars (MSPs) is very sensitive to the gravitational bending of light effect, which in turn is a strong function of the neutron star compactness, *M/R* (Bogdanov, 2013). The dramatic increase in sensitivity enabled by *Athena+* will allow us to reach a dozen of MSPs and will permit a very constraining ~5% measurement in a modest exposure time especially when combined with an improved mass measurement from precision radio timing and constraints on the system geometry from radio and γ-ray pulse modelling (see Fig. 3). For accreting NSs, a similar method can be applied to the strong X-ray pulsations detected during many thermonuclear X-ray bursts. Moreover, the simultaneous measurement of the broad iron line and of the upper QPO frequency offers another efficient mean to measure mass and radius (Cackett et al., 2010, see Fig. 4).

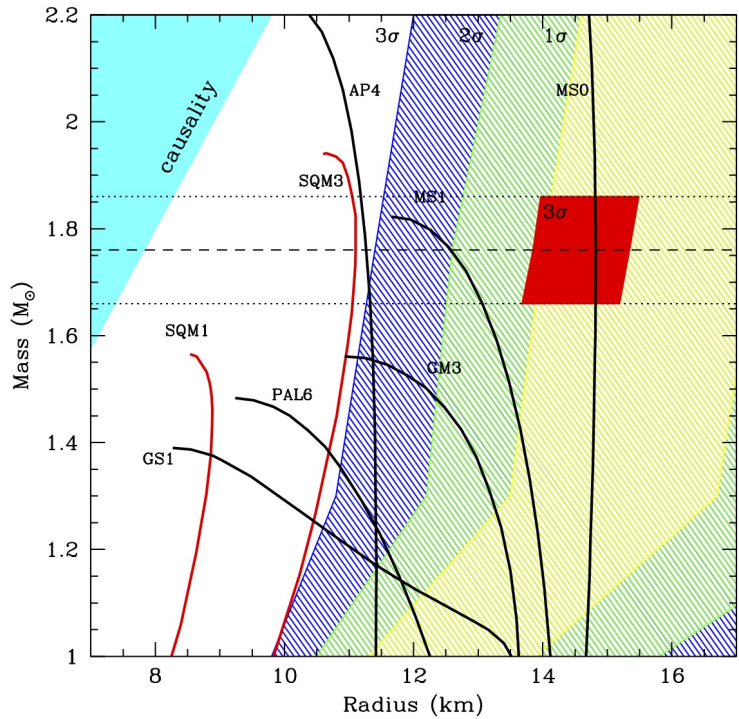

Figure 3: Constraints on the equation of state derived from the modeling of the X-ray light curve of the millisecond pulsar PSR J0437-4715. The mass-radius plane for neutron stars shows the 1, 2, and 3σ confidence contours (yellow, green and blue hatched regions, respectively) based on a 130 ks *XMM-Newton* EPIC pn observation. The solid lines are representative theoretical model tracks (from Lattimer & Prakash, 2001). The horizontal lines show an assumed 1.76 $M_{sun}$ pulsar mass measurement from radio timing and the associated 10% measurement uncertainties (dotted lines). The red region is the expected 3σ constraint from *Athena+* in a 130 ks WFI observation, taking into account uncertainties on the parameters of the magnetic field configuration.

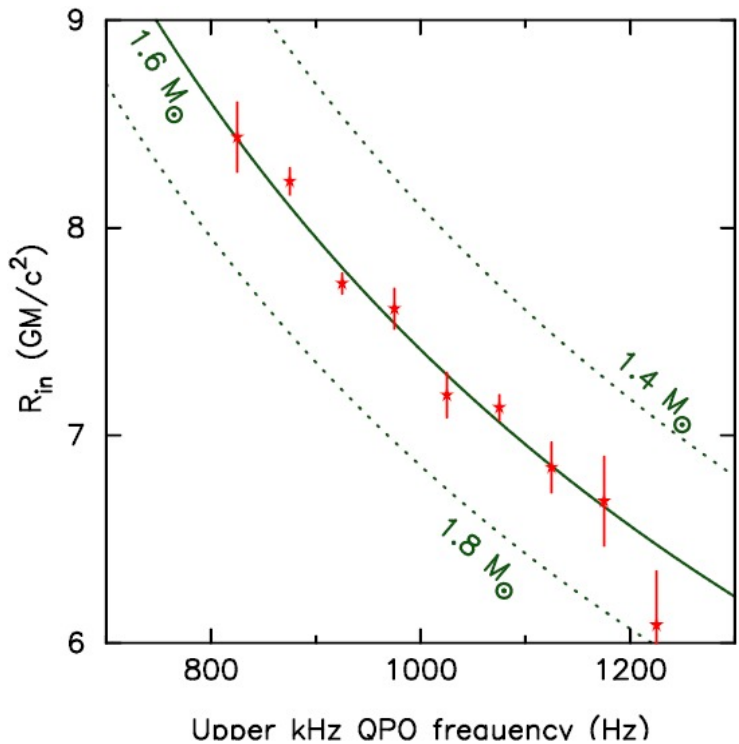

Figure 4. If the upper kHz QPO frequency is associated with the inner edge of the accretion disk, then changes in the frequency will track changes in the inner accretion disk radius, $r_{in}$, which can also be measured through the broad Fe K emission line. The relation between the upper kHz QPO frequency and $r_{in}$ is just set by the neutron star mass. Hence, by simultaneously measuring $r_{in}$ from the Fe K line and the upper kHz QPO frequency from the lightcurve we can measure the neutron star mass. Here we simulated a 100 ks *Athena+* observation of a typical bright neutron star low-mass X-ray binary (LMXB) based on the frequency evolution seen in 4U 1608-52. Splitting the 100 ks observation into 1 ks segments, we measure $r_{in}$ from the spectrum. The red stars (and errors) indicate the measured values binned in frequency, showing we recover the input neutron star mass of 1.6 $M_{sun}$

The detection of gravitationally red-shifted line signatures from the neutron star atmosphere would directly yield the value of $M/R$. Absorption lines from high-$Z$ elements are expected in the X-ray spectra of thermally emitting NSs with residual accretion onto the surface and have been predicted in type I X-ray bursts (Chang et al., 2005). The profile of the line is distorted by contributions from magnetic (Zeeman or Paschen-Back) splitting by the star's magnetic field, longitudinal and transverse Doppler shifts, special relativistic beaming, gravitational redshifts, light bending, and frame dragging. Thanks to its improved effective area and spectral resolution, spectroscopic observations of X-ray bursts with the high spectral resolution of the X-IFU of slowly rotating neutron stars and with the moderate spectral resolution of the WFI for rapidly rotating neutron stars will enable us to detect weak absorption features. Narrow absorption lines have already been detected in the stacked grating X-ray spectra of two of the thermally emitting and slowly rotating isolated neutron stars (Hohle et al., 2012). At present, the origin of these lines either from the NS surface or from circumstellar or interstellar origin remains unknown. The X-IFU will be the first instrument able to discover fainter lines that will constrain the redshift of the emitting region and accurately measure their profile within <10 ks from which M/R can be estimated if the lines originate from the NS atmosphere.

## 4.2. Emission mechanisms in rotation powered neutron stars

Pulsars are unique laboratories as neutron stars are the most compact objects accessible by direct astrophysical observations. Their complex emission mechanisms are far from being understood (Becker 2009). For instance, although it is now clear that γ-rays are emitted in the outer magnetosphere, the regions where X-ray and optical photons are created are not yet established, nor is understood how the corresponding emission processes are related one to each other. Sensitive multi-wavelength spectral and timing observations (e.g., using *Athena+*, the e-ELT and SKA) may solve this issue. X-ray pulse phase spectroscopy with the WFI is also crucial to separate magnetospheric emission from the thermal component emitted by polar regions. Likewise, in magnetars, the weak proton cyclotron line diluted by vacuum polarisation effects will be detected at >10σ in 50 ks, which will give a direct measurement of the debated poloidal field strength. The large field of view and high effective area of *Athena+* will allow for deep searches for pulsar-wind nebulae, which are thought to be powered by the relativistic pulsar wind. Such observations are crucial to study the pulsar's energy loss mechanisms since only a tiny fraction of it (<10%) goes to power the pulsar multi-wavelength emission. More generally *Athena+* will offer new astrophysical diagnostics to investigate how the intriguing different manifestations of neutron stars are related one to each other.

## 5. X-RAY EMITTING WHITE DWARFS

A further natural laboratory in which to study the end stages of stellar evolution are accreting white dwarfs (WDs). X-ray observations of such systems provide unique plasma diagnostics of the disk-WD boundary layer in normal cataclysmic variables (CVs), and the magnetically-channelled accretion flows in (intermediate) polars, where the accreting white dwarf has a very strong intrinsic magnetic field. Like X-ray pulsars, in magnetic CVs most of the X-ray emission is produced in shocked regions – "accretion columns" – close to the magnetic poles of the white dwarf. For such systems in particular, X-IFU will bring, for the first time, sufficient spectral and temporal resolution to make it possible to make dynamical measurements of the accretion flows through emission line shifts. Coupled with the geometrical constraints on the accretion flow already available, for example from the highly-structured rotational light curves, this dynamical information will permit the 3-dimensional reconstruction of the flows, allowing an unprecedented detailed view of the complex interplay between the streaming accreted matter and the strong magnetic field of the accreting white dwarf. *Athena+* is also sensitive enough to allow routinely to measure the gravitational redshifted iron Kα line from the surface of the white dwarf. From such measurements it is possible to determine directly the mass of the white dwarf.

In non-magnetic accreting white dwarfs, the boundary layer between the accretion flow and the dwarf's surface has a temperature around 10 keV. During Dwarf Nova outbursts this hard component disappears as the emission shifts to the EUV, although a distinct soft (~1 keV) X-ray component may appear (Wheatley & Mauche, 2005), possibly due to the same magnetohydrodynamic processes that are responsible for the hard state spectra of microquasars (Wheatley, 2006).

Some accreting white dwarfs show multiple thermonuclear outbursts. Of these, recurrent Novae (RN) have likely very massive WDs close to the Chandrasekhar limit (Yaron et al. 2005). A large population of RNs has been predicted but not yet observed (Schaefer 2010). After the initial explosion, both recurrent and classical Novae undergo a Super Soft X-ray Phase, which may last months to years. Quasi-steady Super-Soft Sources (SSS) have WDs accreting at high rate (~$10^{-7}$ $M_{sun}$ yr$^{-1}$) with non-explosive H-burning. Hence, the SSS status is a strong indication the WD is massive and grows in mass. Because of its high sensitivity at low energies, *Athena+* observations will allow us to measure the temperature, mass, and chemical composition of these objects, and through observations of SSS in the Local Group will provide, for the first time, a large sample of well understood SN Ia progenitors.